\documentclass[conference]{IEEEtran}
\IEEEoverridecommandlockouts
\usepackage{cite}
\usepackage{amsmath,amssymb,amsfonts,gensymb}
\usepackage{algorithmic}
\usepackage[final]{graphicx} %included the word final in order to resolve imaging issues...
\usepackage{mathtools, cuted}
\usepackage{textcomp}
\usepackage{xcolor}
\usepackage[export]{adjustbox} %adjusts the table formatting
\usepackage{caption} % puts the table number and title on the same line
\usepackage{subcaption} %auto assign the sub float figure caption numbers
\usepackage{multirow}
\usepackage [latin1]{inputenc}
\renewcommand{\baselinestretch}{0.94}
 
\begin{document}

\title{Propagation Measurements and Coverage Analysis for mmWave and Sub-THz Frequency Bands with Transparent Reflectors\\
}

\author{Ashwini P. Ganesh$^*$,  Wahab Khawaja$^*$, Ozgur Ozdemir$^*$, \.{I}smail G\"{u}ven\c{c}$^*$,
Hiroyuki Nomoto$^\dagger$, and Yasuaki Ide$^\dagger$\\
$^*$Department of Electrical and Computer Engineering, NC State University, Raleigh, NC 27606\\
$^\dagger$Research \& Development Institute, High-Performance Plastics Company, Sekisui Chemical CO., LTD.\\
%https://www.overleaf.com/project/6339ca3e3856794c90decb85
e-mail:~\{apondey, wkhawaja, oozdemi, iguvenc\}@ncsu.edu,~~\{hiroyuki.nomoto, yasuaki.ide\}@sekisui.com
\thanks{This work has been supported in part by the National Science Foundation (NSF) through the award  CNS-1916766 as well as the industry membership fees from the Broadband Wireless Access Center (BWAC) I/UCRC Center.} 
}
\maketitle

\begin{abstract}
The emerging 5G and future 6G technologies are envisioned to provide higher bandwidths and coverage using millimeter wave (mmWave) and sub-Terahertz (THz) frequency bands. The growing demand for higher data rates using these bands can be addressed by overcoming high path loss, especially for non-line-of-sight (NLOS) scenarios. In this work, we investigate the use of passive transparent reflectors to improve signal coverage in an NLOS indoor scenario. Measurements are conducted to characterize the maximum reflectivity property of the transparent reflector using channel sounder equipment from NI. Flat and curved reflectors, each with a size of 16 inches by 16 inches, are used to study coverage improvements with different reflector shapes and orientations. The measurement results using passive metallic reflectors are also compared with the ray-tracing-based simulations, to further corroborate our inferences. The analysis reveals that the transparent reflector outperforms the metal reflector and increases the radio propagation coverage in all three frequencies of interest: 28~GHz, 39~GHz, and 120~GHz. Using transparent reflectors, there is an  increase in peak received power that is greater than 5~dB for certain scenarios compared to metallic reflectors when used in flat mode, and greater than 3~dB  when used in curved (convex) mode. 
%The performance of the transparent reflector can be best seen at 120~GHz which suggests that these reflectors can be used to increase the wave-guiding at higher sub-THz frequencies. At these frequencies, we have fewer reflections in LOS directional channels due to narrower beam antennas at both ends of the link. 

\end{abstract}


\section{Introduction}

% 1 paragraph motivation on coverage challenges for mmWave, and how reflectors can help

One of the foremost challenges in millimeter wave (mmWave) and sub-THz networks is their ability to continuously maintain widespread and resilient wireless coverage~\cite{7996974,erden2021emr,aykin2019multi}. This requires sustaining a line-of-sight (LOS) or at least a strong first-order reflected non-LOS (NLOS) path to form a stable link for radio communications. The availability of an NLOS link depends on the reflection profile of a scatterer, which is characterized by the scatterer's material and the frequency of operation. A conventional approach to providing a stable NLOS link is by the deployment of multiple access points or active repeaters~\cite{ozdemir202028} and relays~\cite{biswas2015performance}. Strategical placements of reflectors in indoor and outdoor  environments were studied in~\cite{9840381,anjinappa2020base} using analytical derivations, which show the strong dependency of wireless coverage on the location and the size of the reflectors.

Recent studies on intelligent reflecting surfaces (IRS) show that the environment can be controlled to steer electromagnetic waves in desired directions, which makes the network less prone to disruptions such as blockage~\cite{liaskos2018new, 9771079,ozdogan2019intelligent,9140329}. However, the use of IRSs requires control circuitry and requires a high price and computation cost.  
Other solutions proposed to improve the wireless coverage at mmWave and sub-THz bands include increasing the transmit power or deploying highly-sensitive receivers. Due to rules on peak transmit power by regulatory bodies and the high cost of highly sensitive receivers, one of the promising alternatives to improve wireless coverage is by using passive reflectors. The deployment of passive reflectors with good reflection characteristics can be used to our advantage to replace the multiple access points and repeaters. Earlier simulation and measurement studies, such as~\cite{khawaja2020coverage} have shown that objects such as metal act as a perfect reflector, enabling strong reflections for directional NLOS communication. Thus, strategic deployment of passive metallic reflectors can create a favorable propagation environment by introducing new multi-path components (MPCs) in the channel and increasing the overall spatial diversity of the MPCs. Metals, however, may be undesirable to be placed unconditionally and abundantly as reflectors due to cosmetic and other reasons, especially in indoor environments. 

In places where the deployment of metallic reflectors is not reasonable, transparent reflectors can offer a sustainable, easy-to-deploy alternative to increase the network coverage without altering the appearance of the surface. Penetration properties of these transparent reflectors in an indoor and an open-door environment were studied in our recent work~\cite{anjinappa2022indoor}. The experiment result showed better reflection performance, and improvement in coverage by preserving the radio waves within the environment in presence of passive transparent reflectors. Amongst the different kinds of reconfigurable intelligent surfaces (RISs) being researched currently~\cite{9140329}, our film under test is electromagnetic-based (EM) based RIS. The film withholds the capacity to preserve the impinging radio waves and passively reflect the first-order derivative for enhancing the network coverage. These reflectors from Sekisui, as tested initially in~\cite{anjinappa2022indoor}, provide a strong NLOS link in a wireless network and have a penetration loss comparable to metal. Hence, this film might be an economical and sustainable alternative to other conventional methods of increasing the received signal strength and indoor coverage. 
%The conventional methods are undoubtedly effective but are subjected to higher infrastructure and deployment costs. 

Another distinctive advantage of this film is its non-planar structure and free-form conformality that leads to easy deployment, high reusability, and durability. They reflect all the incident rays enabling them to reach shielded, dense areas eliminating any blind spots. These transparent reflectors can be deployed on any structure without altering the appearance of the surface that it is mounted on. %The diffusion characteristics of the reflector film also enable a wide frequency coverage useful for high-throughput 5G and 6G communications in mmWave and sub-THz bands.

\graphicspath{ {./images/} }
\begin{figure}[t]
\centering
\includegraphics[scale=0.75, width=.45\textwidth,height=.35\textwidth]{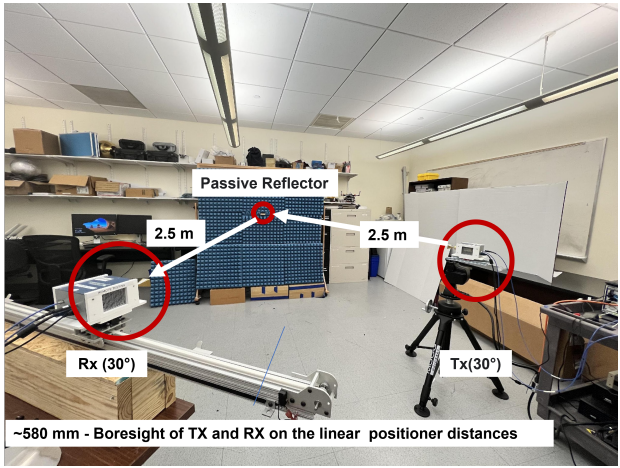}
\caption{Measurement setup for evaluating reflector characteristics using a channel sounder that operates at multiple mmWave and sub-THz frequencies.}
\label{fig:Testscenario} 
\end{figure}

In this work, we characterize the propagation characteristics of passive transparent reflectors in a controlled indoor environment. %We conducted our experiments in an intelligent radio environment, motivated by the high importance of a controllable environment to improve signal coverage. 
Our measurement setup that uses a channel sounder system from NI is shown in Fig.~\ref{fig:Testscenario}. A panel made of broadband AEP-04 pyramidal absorbers from MVG~\cite{Pyramid_Absorber} is used to eliminate other second-order reflections. Our main contributions can be listed as follows: 
\begin{itemize}
\item We conducted measurements with different shapes of metallic and transparent reflector and characterized their coverage across an NLOS area using a linear positioner (controlled by LabVIEW) that carries the receiver; 
\item We compared the coverage performance of metallic and transparent reflectors at 28~GHz, 39~GHz, and 120~GHz, at various different receiver locations with respect to the reflector's location; 
\item We developed ray tracing simulations to compare the coverage of flat and curved metallic reflectors at different locations, and compared the ray-tracing result with measurements which indicate a close match.  
\end{itemize}

The rest of this paper is organized as follows. In Section~\ref{Sec:2}, we describe details of our measurement scenario, while Section~\ref{Sec:3} describes our model for ray tracing simulations with flat and curved metallic reflectors. Section~\ref{Sec:4} presents our measurement and simulation results with reflectors and the last section concludes the paper. 

\graphicspath{ {./images/} }
\begin{figure}[t]
\centering
    \subfloat[Convex metal reflector.]
    {\includegraphics[width=.23\textwidth,height=.18\textwidth]{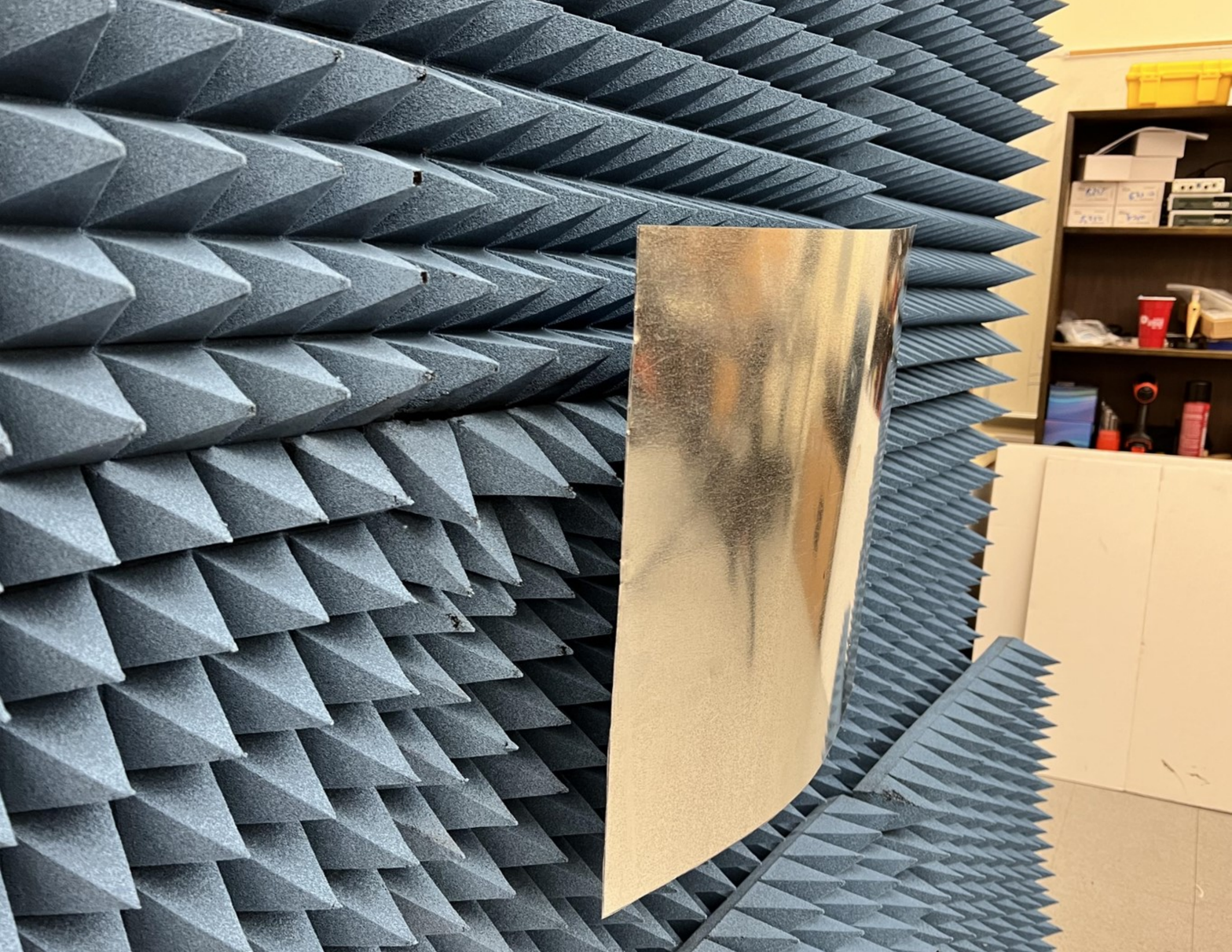}}\quad
    \subfloat[Convex transparent reflector.]
    {\includegraphics[width=.23\textwidth,height=.18\textwidth]{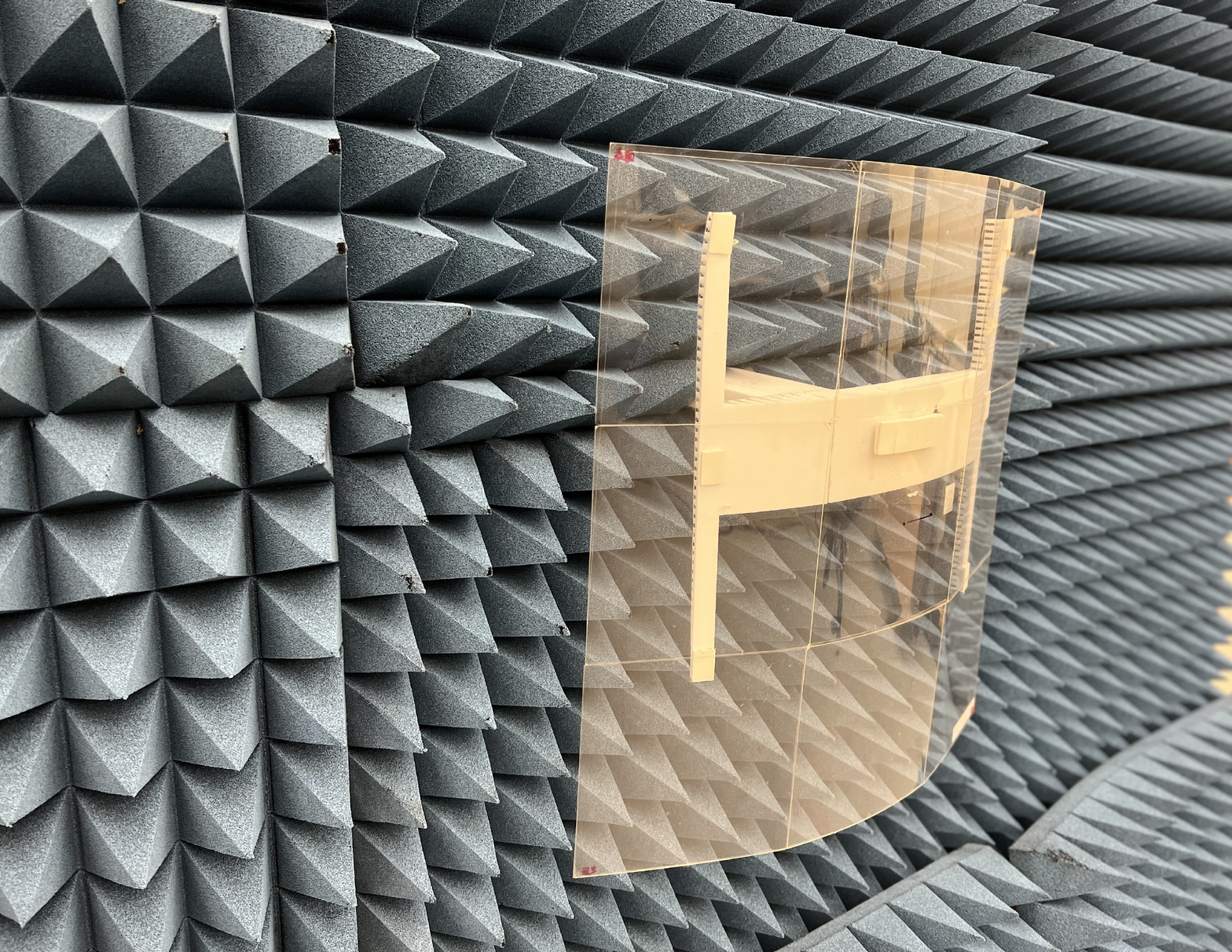}}\quad
    \subfloat[Flat metal reflector.]
    {\includegraphics[width=.23\textwidth,height=.18\textwidth]{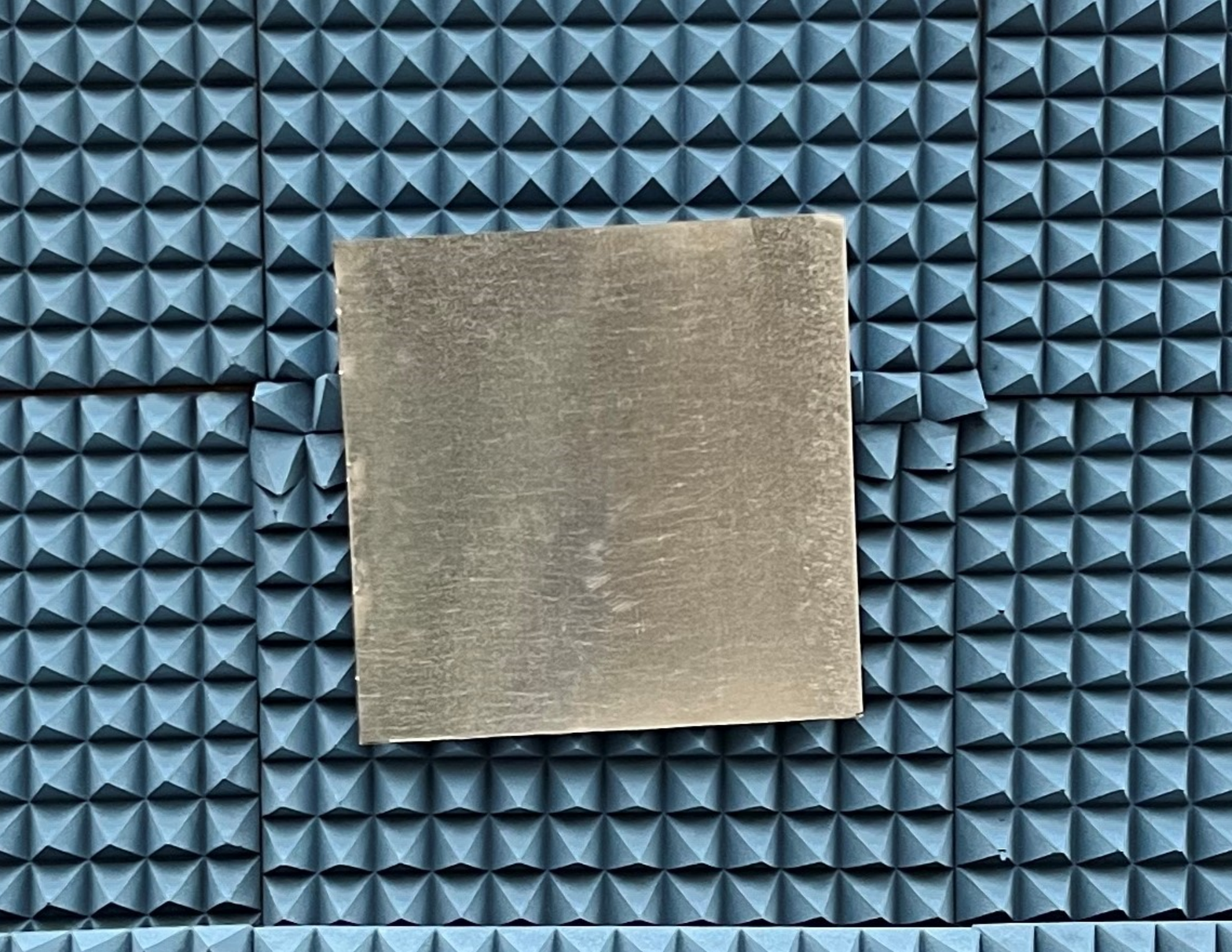}}\quad
    \subfloat[Flat transparent reflector.]
    {\includegraphics[width=.23\textwidth,height=.18\textwidth]{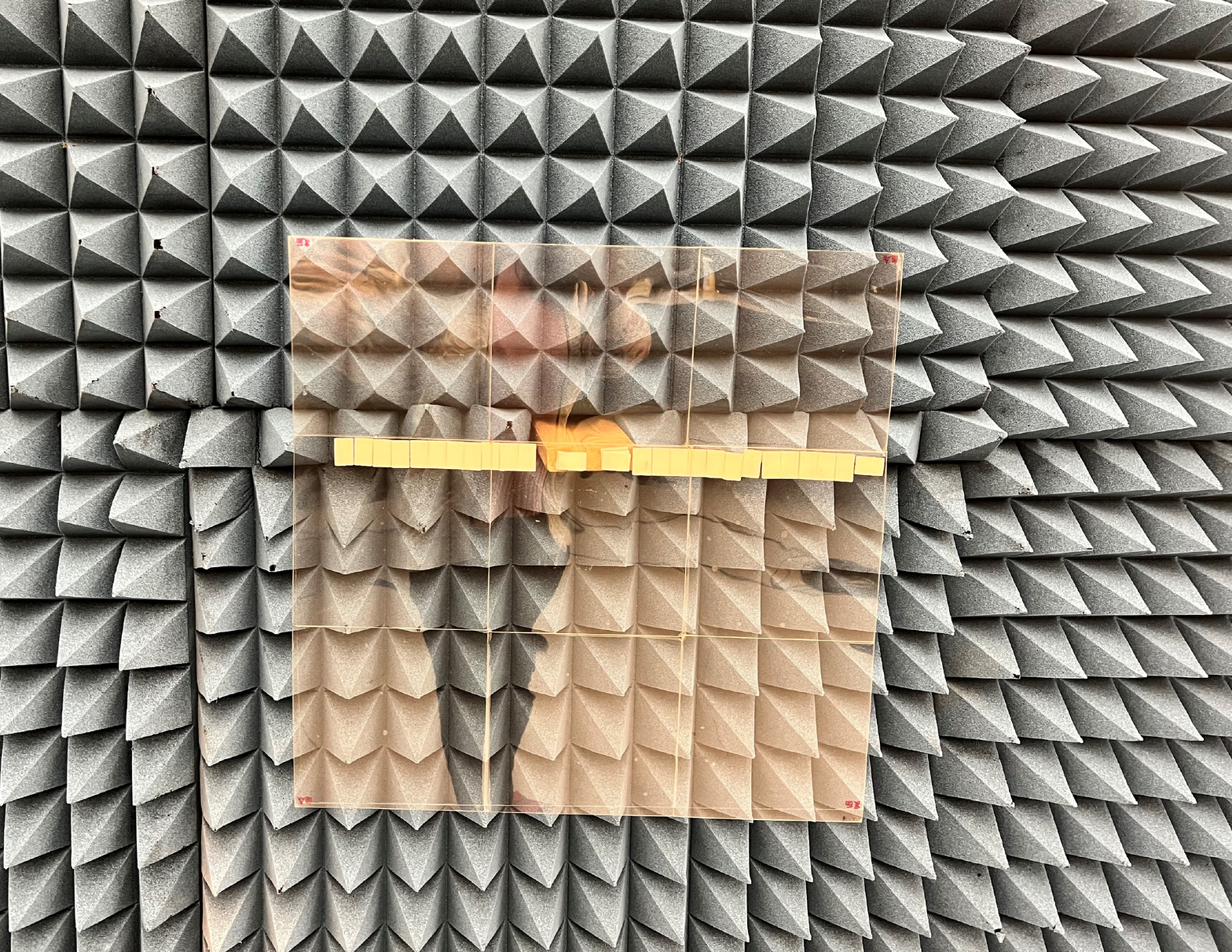}}
\caption{Flat and curved passive reflectors of size 16$"$x16$"$.}
\label{fig:reflectorform} \vspace{-1mm}
\end{figure}

\section{Reflector Measurements Scenario and Setup}\label{Sec:2}
\begin{table}[t]
    \centering
    \renewcommand{\arraystretch}{1.2}
   \begin{tabular}{|c|c|c|c|}\hline
       \textbf{Frequency (GHz)}  & 28& 39& 120 \\ \hline
       \textbf{IF (GHz)}& 10.56 & 10.56 & 12.6 \\
       \hline
       \textbf{Bandwidth (GHz)} &1.5 &1.5  &1.5\\ 
       \hline
       \textbf{Transmitted power (dBm)} & -10 & -10 & 10\\
       \hline
       \textbf{TX antenna gain (dBi)} & 17 &20 &21\\
       \hline
       \textbf{RX antenna gain (dBi)} & 17 &20 &21\\ \hline
    \end{tabular}
    \caption{Measurement parameters for the channel sounder.}
    \label{tab:measurement_param}\vspace{-4mm}
\end{table}

The measurements were conducted in the time domain using an impulse channel sounder that is built  using NI's PXI platform~\cite{NI_Webpage} that is controlled by LabVIEW. The sounder operates at an intermediate frequency of 10.56~GHz for the central frequencies of 28~GHz and 39~GHz, and an intermediate frequency of 12.6~GHz for the central frequency of 120~GHz. The system up and down converts the intermediate frequency using the PXIe-3620 module along with a PXIe-7902 FPGA. The measurement parameters of each frequency of operation are provided in Table~\ref{tab:measurement_param}. More information on the design parameters of the channel-sounding chassis can be found in our past work~\cite{anjinappa2022indoor}. 

In this work, we evaluate the performance of our propagation channel by measuring the maximum reflected power using metal and transparent reflectors placed in an indoor environment. The evaluation is conducted using the power delay profile (PDP) of the channel. The transmitter (TX) and the receiver (RX) antenna radio heads were both placed on a steerable gimbal and a linear positioner, respectively, as highlighted in Fig.~\ref{fig:Testscenario}. The antennas used are directional horn antennas that have a 17~dBi, 20~dBi, 21~dBi TX and RX antenna gain at 28~GHz, 39~GHz, and 120~GHz, respectively. At 28~GHz, the antennas have an aperture of 26$^{\circ}$ at 3~dB beam-width (HPBW) in the E-plane and 24$^{\circ}$ in H-plane. The antenna system has a maximum linear dimension of 4.072~cm with a far field of 30.97~cm. At 39 GHz, the aperture is 15$^{\circ}$ in the E-plane and 16$^{\circ}$ in the H-plane, respectively. At 120 GHz, the horn antennas have an aperture of 13$^{\circ}$ in both E-plane and H-plane, respectively. More details on the sounder system can be found at~\cite{kairui}. 

The RX is placed on a tGlide modular linear positioner that travels a distance of 1.8~meters for our measurements. The linear positioner uses a timing belt drive for a custom number of measurements~\cite{Linear_positioner}. The maximum reflected power received by the RX antenna is measured when moving along the positioner with a resolution of 0.1 cm and 1800 measurements at each point. The reflector is placed at a distance of 2.5~m angled at 30$^{\circ}$ perpendicular to the TX and RX. A $16"\times 16"$ reflector is placed on an absorber panel~\cite{Pyramid_Absorber} to eliminate the second-order reflections from other sources in the background. The reflector is placed in between the absorber panel along the elevation axes of the transmitter and receiver radio heads. The reflector helps to make at least one strong first-order non-LOS (NLOS) reflection path to complete the radio link. The flat and curved shape reflectors used in the measurement setup are shown in Fig.~\ref{fig:reflectorform}. The reflector, TX, and RX antenna placement is depicted in Fig.~\ref{fig:Testscenario}. A time alignment was performed before each set of measurements to make sure the TX and RX are calibrated. To limit the influence of frequency and channel response, the measurements were carried out back-to-back.

\section{Ray-Based Simulations for Metal Reflectors}\label{Sec:3}
\begin{figure}[t] 
    \centering
		\includegraphics[width=0.7\columnwidth]{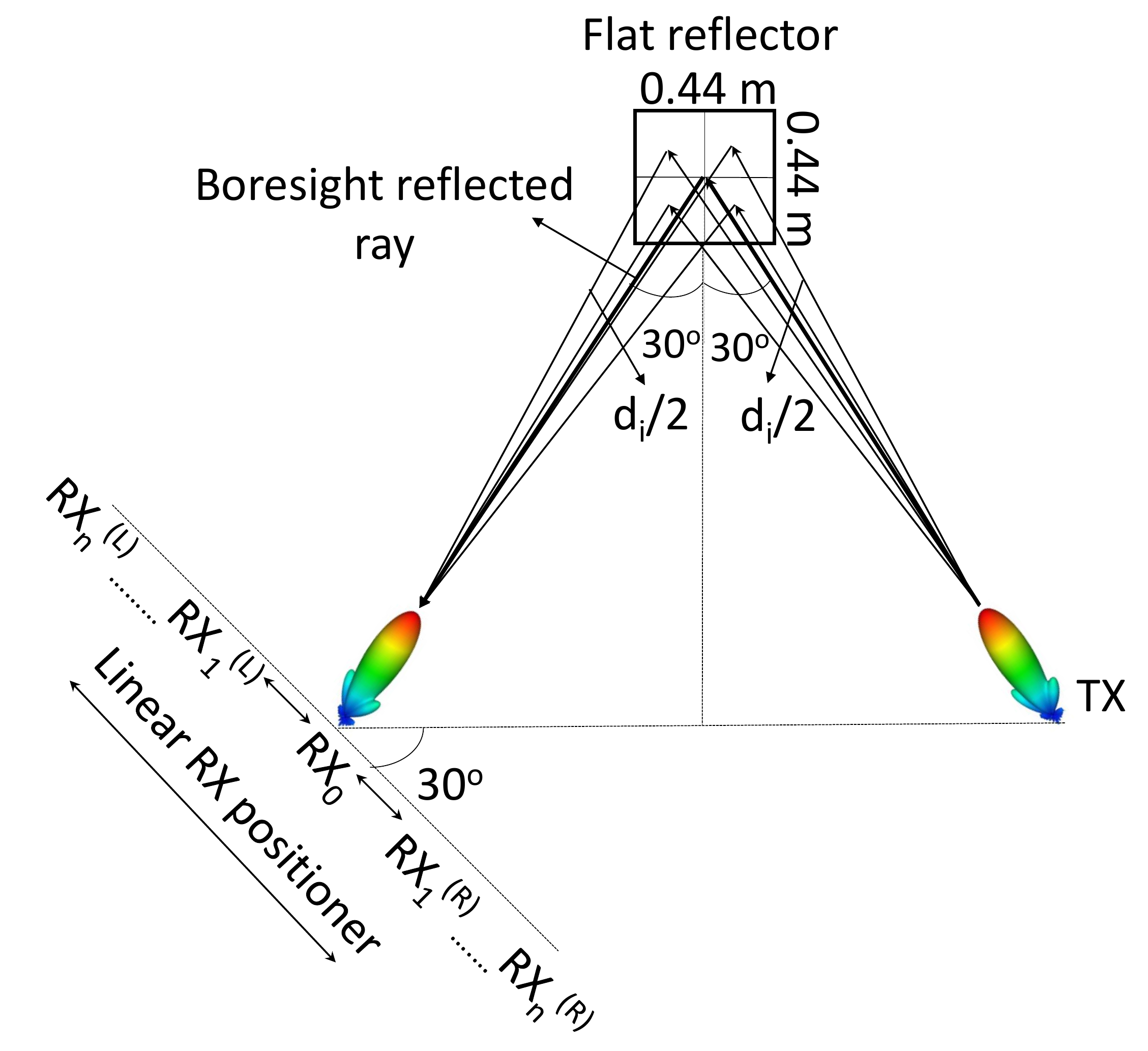}
	   \caption{A four-facet flat reflector scenario for the ray-based simulations. The RX positions on the LHS and RHS are represented as RX$_{k}^{(\rm L)}$ and RX$_{k}^{(\rm R)}$, respectively, and $k=1,2,\hdots,n$. } \label{Fig:simulation_scenario_flat}\vspace{-2mm}
\end{figure}

In addition to the channel measurements, we also carried out ray-based simulations for the metallic reflectors, using both flat and convex reflector scenarios. 
The ray-based simulation setup for the flat reflector is shown in Fig.~\ref{Fig:simulation_scenario_flat}. We divided the metal reflector into a number of facets. We assume that a single ray is incident at the center of every facet and subsequently reflected towards the RX. Each ray is assigned an antenna gain based on the azimuth and elevation angles at the TX and RX. Each ray also covers a given distance from the TX toward the RX. The received power due to rays from the reflector center and $N^{(\rm f)}$ facets at the center RX position, RX$_0$ is given as  
\begin{align}
\small
    P_{\rm R}^{(\rm RX_0)} &= \nonumber  
    %\frac{N^{(\rm f)}}{\sqrt{N^{(\rm f)}}}
    \sqrt{N^{(\rm f)}} %IG: Not sure why writing it separately as above, updated  
    \sum_{i=1}^{N^{(\rm f)}+1}\frac{P_{\rm T}}{(4\pi d_i)^2}\nonumber\\
    &\times\sqrt{G^{(\rm Az)}_{\rm T}\Big(\theta_i^{(\rm cent,TX)}\Big)G^{(\rm El)}_{\rm T}\Big(\phi_i^{(\rm cent,TX)}\Big)} \nonumber \\ 
    &\times\sqrt{G^{(\rm Az)}_{\rm R}\Big(\theta_i^{(\rm cent,RX)}\Big)G^{(\rm El)}_{\rm R}\Big(\phi_i^{(\rm cent,RX)}\Big)} \nonumber \\ 
    & \times\lambda^2 \exp{\bigg(-j2\pi\frac{(d_i-d_{\rm ref})}{\lambda}}\bigg)\alpha^{(\rm flt)}~, 
        \label{Eq:RX_pwr_flat}
\normalsize
\end{align}
where $P_{\rm T}$ is the transmit power, $G^{(\rm Az)}_{\rm T}\big(\theta_i^{(\rm cent,TX)}\big)$ and $G^{(\rm El)}_{\rm T}\big(\phi_i^{(\rm cent,TX)}\big)$ are the TX antenna gains in the azimuth and elevation planes, respectively, of the $i^{\rm th}$ ray, and $\theta_i^{(\rm cent,TX)}$ and $\phi_i^{(\rm cent,TX)}$, represent the azimuth and elevation angles of the $i^{\rm th}$ ray, respectively, at the TX. 

The RX antenna gain of $i^{\rm th}$ ray in the azimuth and elevation plane is represented as $G^{(\rm Az)}_{\rm R}\big(\theta_i^{(\rm cent,RX)}\big)$ and $G^{(\rm El)}_{\rm R}\big(\phi_i^{(\rm cent,RX)}\big)$, respectively, and $\theta_i^{(\rm cent,RX)}$ and $\phi_i^{(\rm cent,RX)}$ represent the azimuth and elevation angles at the RX, respectively. The distance traveled by the $i^{\rm th}$ ray from the TX to the RX through the reflector is given as $d_i$ and the phase of each ray is represented by $\exp{\Big(-j2\pi\frac{(d_i-d_{\rm ref})}{\lambda}}\Big)$, where $d_{\rm ref}$ is the reference distance between the boresight of the TX and RX antennas through the center of the reflector and $\alpha^{(\rm flt)}$ represents the attenuation due to the difference in the area of the electromagnetic energy incident at the reflector compared to the physical area of the reflector. %Similarly, the received power at other RX positions on the linear positioner are obtained.

\begin{figure}[t] 
    \centering
		\includegraphics[width=\columnwidth]{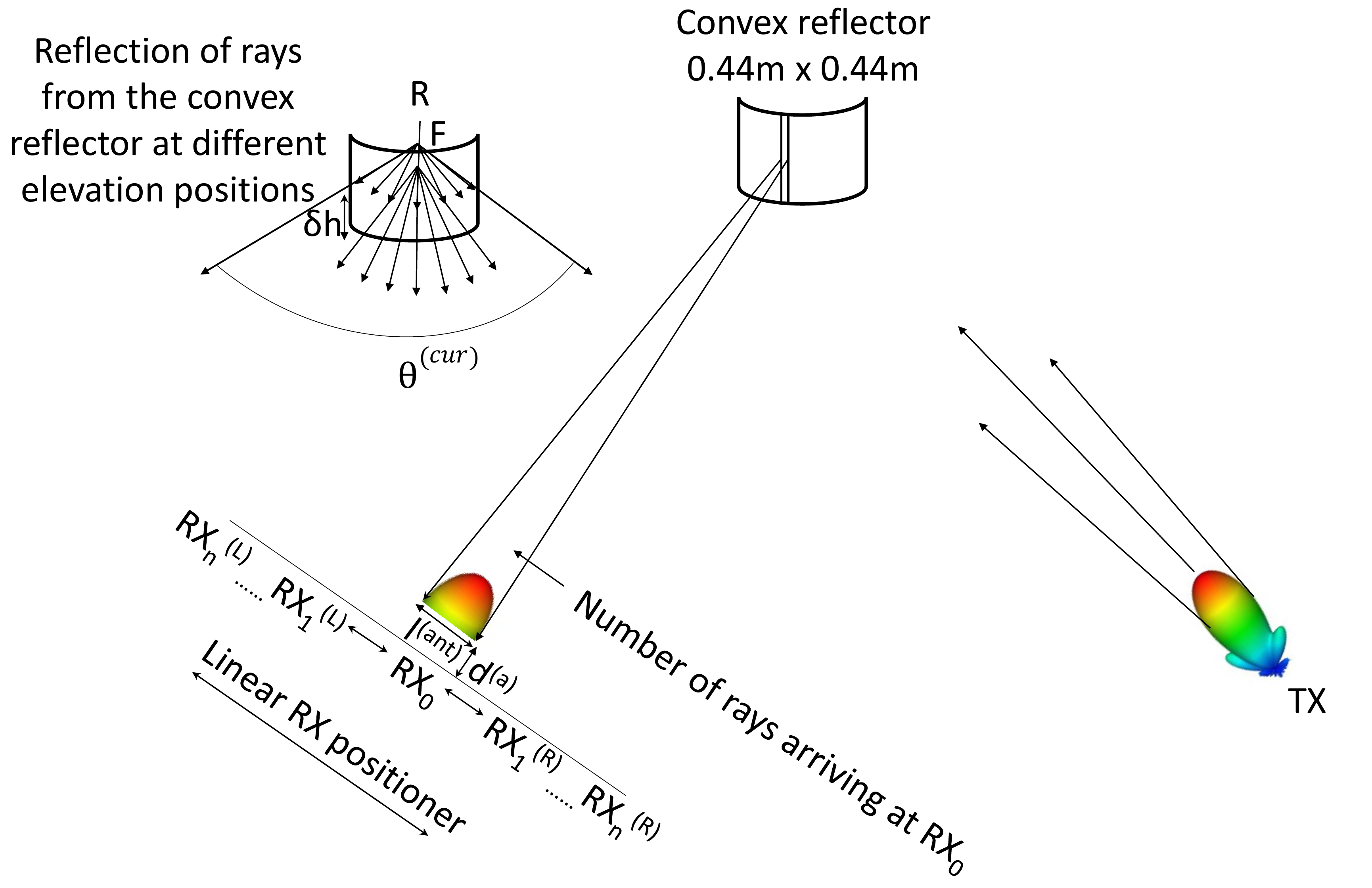}
	   \caption{Ray-based simulation scenario for the convex reflector.} \label{Fig:simulation_scenario_convex}\vspace{-2mm}
\end{figure}

The ray-based simulation scenario for the convex-shaped reflector is shown in Fig.~\ref{Fig:simulation_scenario_convex}. The rays reflected from the convex reflector diverge in different directions in the azimuth plane, with  azimuth divergence angles given by $\theta^{(\rm cur)}$. The diverging rays converge virtually at a focal point at $F=R/2$ virtually beyond the convex reflector and $R$ is the radius of curvature.
Moreover, the convex reflector is divided into different sections of height. The difference in height of any two sections is represented as $\delta h$ as shown in Fig.~\ref{Fig:simulation_scenario_convex}. For a given RX position on the linear positioner, only a finite number of rays can be captured by the antenna of length $l^{(\rm ant)}$ in the azimuth plane at a far field distance $d^{(\rm a)}$. The length $ l^{(\rm ant)}$ depends on the half power beamwidth of the antenna in the azimuth plane. The received power for the convex reflector is given as:
\begin{align}
\small
P_{\rm R}^{(\rm RX_0)} &= \sqrt{N^{(\rm El)}N^{(\rm Az)}}\sum_{i=1}^{N^{(\rm El)}}\sum_{j=1}^{N_i^{(\rm Az)}}\frac{P_{\rm T}}{(4\pi d_{i,j})^2} \nonumber \\ &\times \sqrt{G^{(\rm Az)}_{\rm T}\Big(\theta_{i,j}^{(\rm cent,TX)}\Big)G^{(\rm El)}_{\rm T}\Big(\phi_{i,j}^{(\rm cent,TX)}\Big)} \nonumber \\ &\times \sqrt{G^{(\rm Az)}_{\rm R}\Big(\theta_{i,j}^{(\rm cent,RX)}\Big)G^{(\rm El)}_{\rm R}\Big(\phi_{i,j}^{(\rm cent,RX)}\Big)} \nonumber \\ &\times \lambda^2 \exp{\bigg(-j2\pi\frac{(d_{i,j}-d_{\rm ref})}{\lambda}}\bigg)\alpha^{(\rm cur)},
        \label{Eq:RX_pwr_curve}
\normalsize
\end{align}
where $N^{(\rm El)}$ is the number of height sections considered given as $N^{(\rm El)} = \frac{h}{\delta h}$, where $h$ is the height of the convex reflector, $N_{i}^{(\rm Az)}$ is the number of rays in the azimuth plane at the $i^{\rm th}$ height section and $N^{(\rm Az)} = \frac{l^{(\rm ant)}}{\gamma}$, and $\alpha^{(\rm cur)}$ is the attenuation factor for the convex reflector. %The attenuation factor for the convex reflector is determined by the area of the reflector compared to the area of incident electromagnetic energy and divergence of rays away from a given RX position. 
The attenuation for the convex reflector is larger compared to the flat reflector, $\alpha^{(\rm cur)}<\alpha^{(\rm flt)}$. % given as 
%\begin{align}
 %   &\alpha^{(\rm cur)} = \frac{A^{(\rm cur)}}{A^{(\rm bm)}}\times f\big(\theta^{(\rm cur)}\big), 
  %      \label{Eq:atten_diverge_convex}
%\end{align}
%where $A^{(\rm cur)}$ is the area of the curved reflector, and $f\big(\theta^{(\rm cur)}\big)$ is the reduction factor. $A^{(\rm cur)}f\big(\theta^{(\rm cur)}\big)$ represents the reduced reflector area responsible for the reflection of the rays towards a given RX position as shown in Fig.~\ref{Fig:simulation_scenario_convex}. The reduction factor $f$ depends on the divergence angle $\theta^{(\rm cur)}$ which in turn depends on the radius of curvature $R$ and size of the reflector. Overall, greater the divergence angle, greater the spread of the rays, however, the RX power of the rays are attenuated proportionally by the factor $f\big(\theta^{(\rm cur)}\big)$ for a given RX position. 

\section{Measurement and Simulation Results}\label{Sec:4}
In this section, measurement results for the metallic and Sekisui transparent reflectors are provided. The ray-based simulation results for metallic reflectors are also compared with the measurement results.
After conducting the measurements using channel sounder equipment, we processed the data points obtained in Matlab and averaged them across all multipath reflections in the NLOS environment. 

\subsection{Measurement Results Analysis}
\begin{figure*}[h!]
  \centering
     \begin{subfigure}[b]{0.45\textwidth}
        \centering
        {\subfloat[Frequency: 28~GHz, Shape: Flat]{\includegraphics[width=\textwidth]{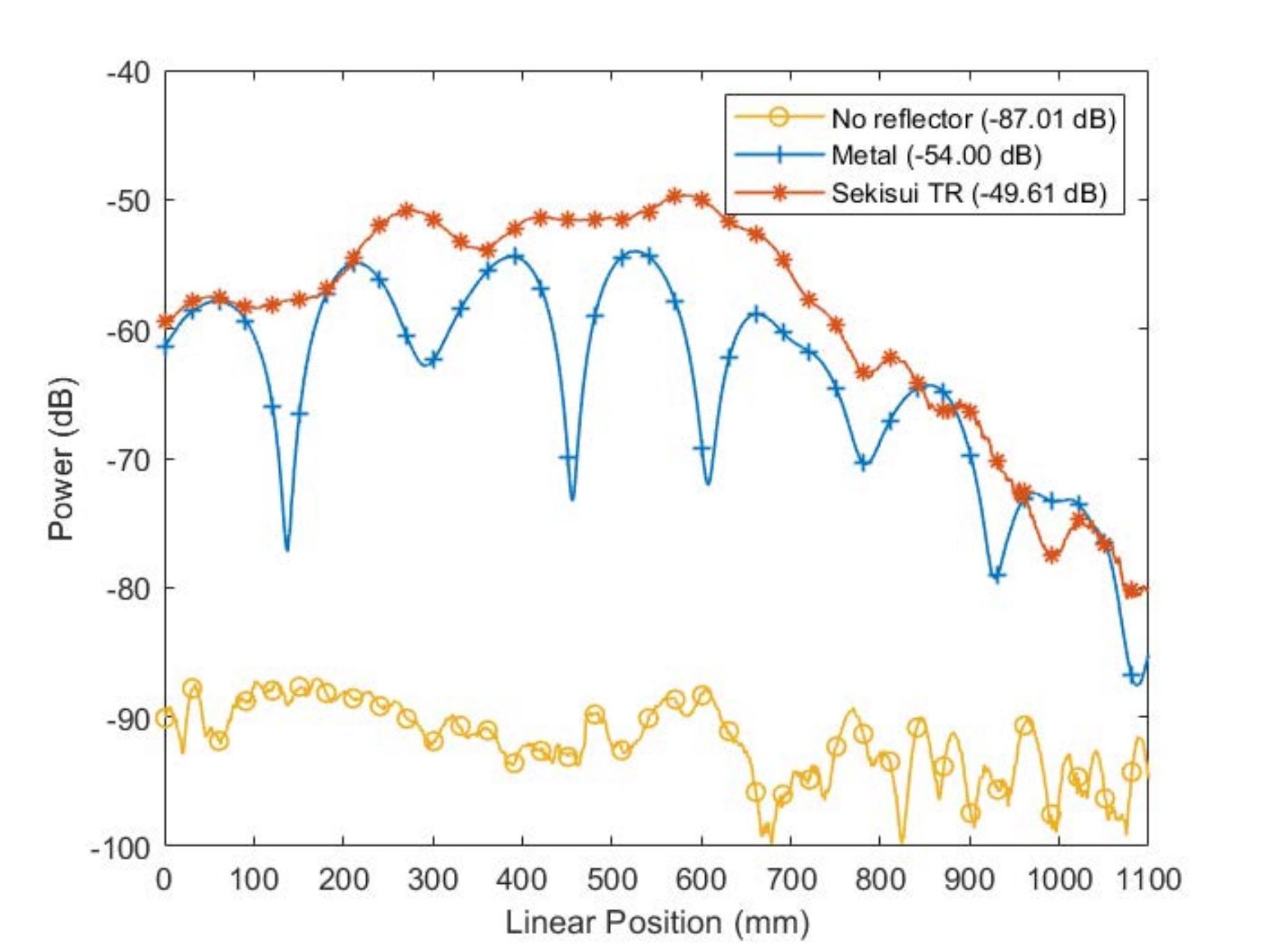}}}
        % \caption{Freq: 28 GHz, Reflector Type: Flat}
    \end{subfigure}
    \begin{subfigure}[b]{0.45\textwidth}
       {\subfloat[Frequency: 28~GHz, Shape: Convex]{\includegraphics[width=\textwidth]{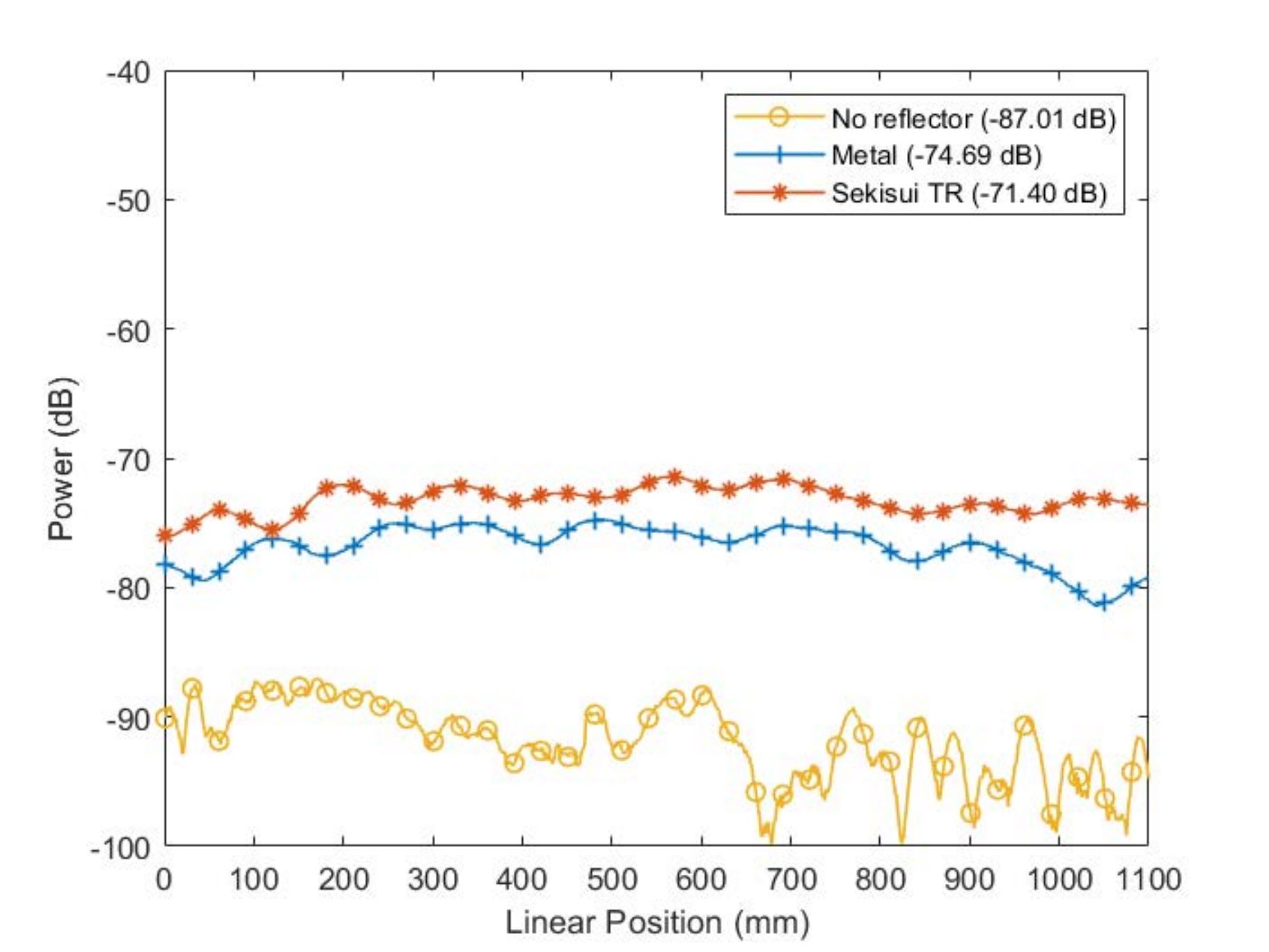}}}
       \centering
    %   \caption{Freq: 28 GHz, Reflector Type: Convex}
    \end{subfigure}
    \\
         \begin{subfigure}[b]{0.45\textwidth}
        \centering
        {\subfloat[Frequency: 39~GHz, Shape: Flat]{\includegraphics[width=\textwidth]{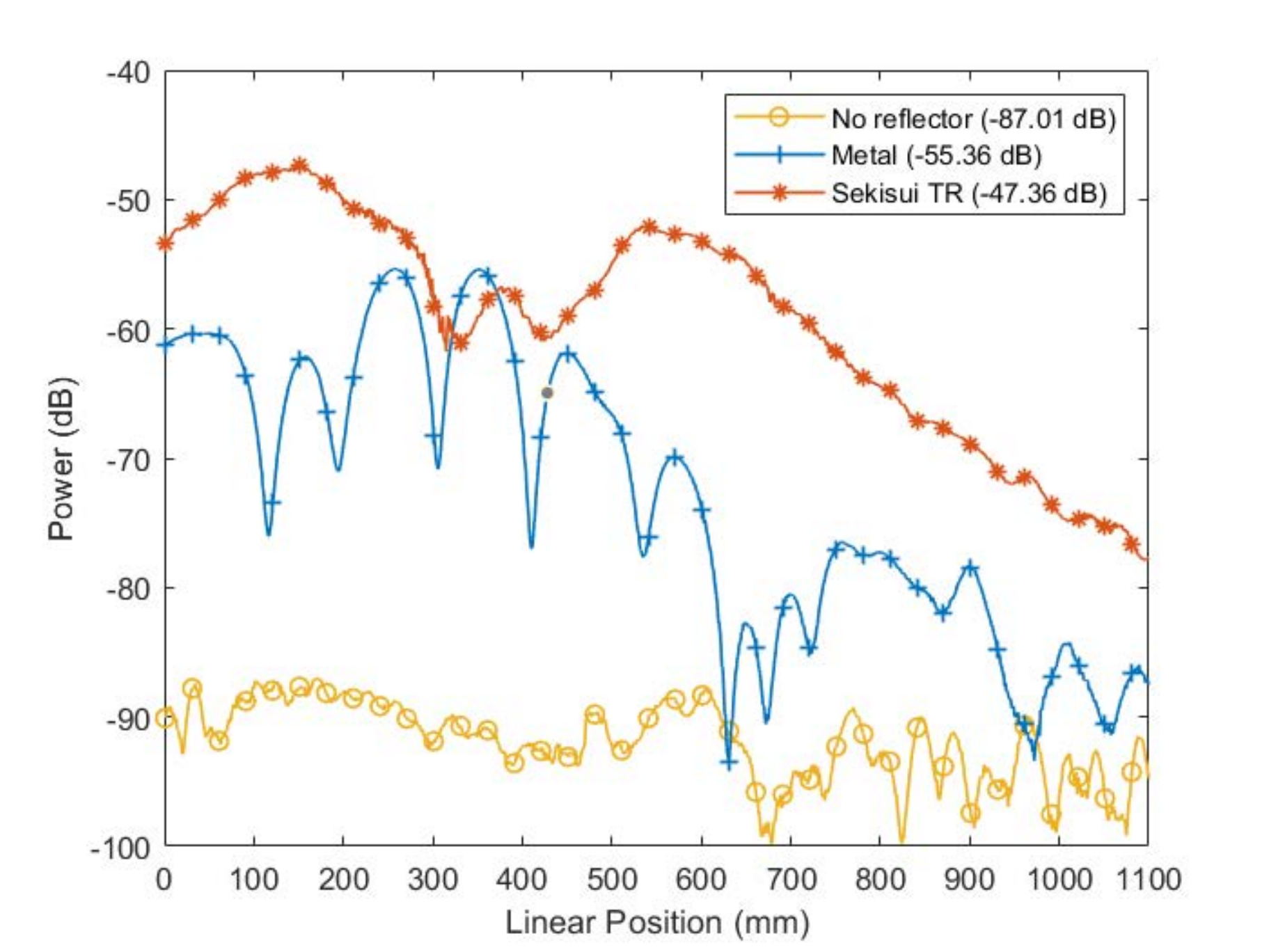}}}
        % \caption{Freq: 39 GHz, Reflector Type: Flat}
    \end{subfigure}
    \begin{subfigure}[b]{0.45\textwidth}
       {\subfloat[Frequency: 39~GHz, Shape: Convex]{\includegraphics[width=\textwidth]{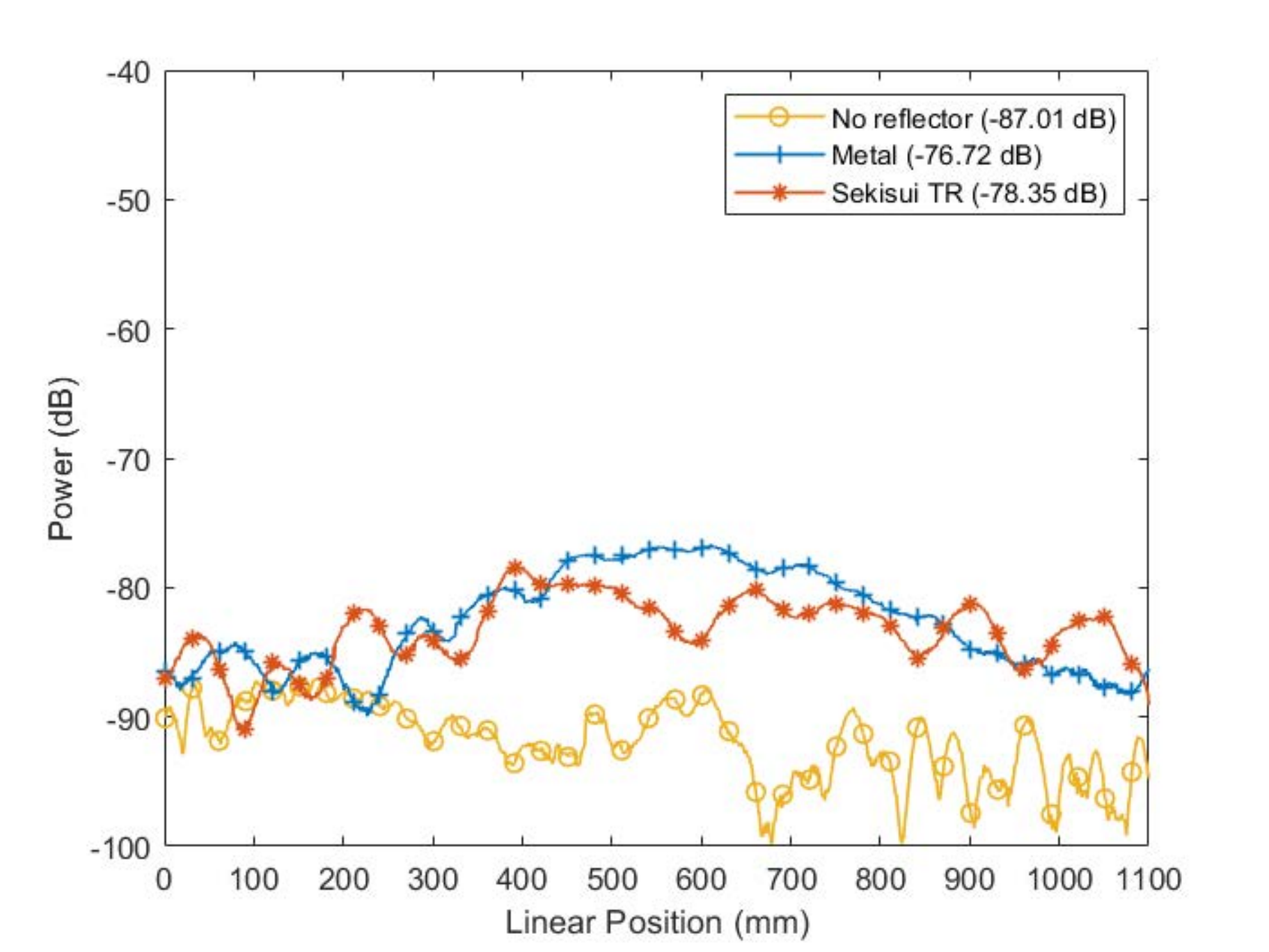}}}
       \centering
    %   \caption{Freq: 39 GHz, Reflector Type: Convex}
    \end{subfigure}
    \\
         \begin{subfigure}[b]{0.45\textwidth}
        \centering
        {\subfloat[Frequency: 120~GHz, Shape: Flat]{\includegraphics[width=\textwidth]{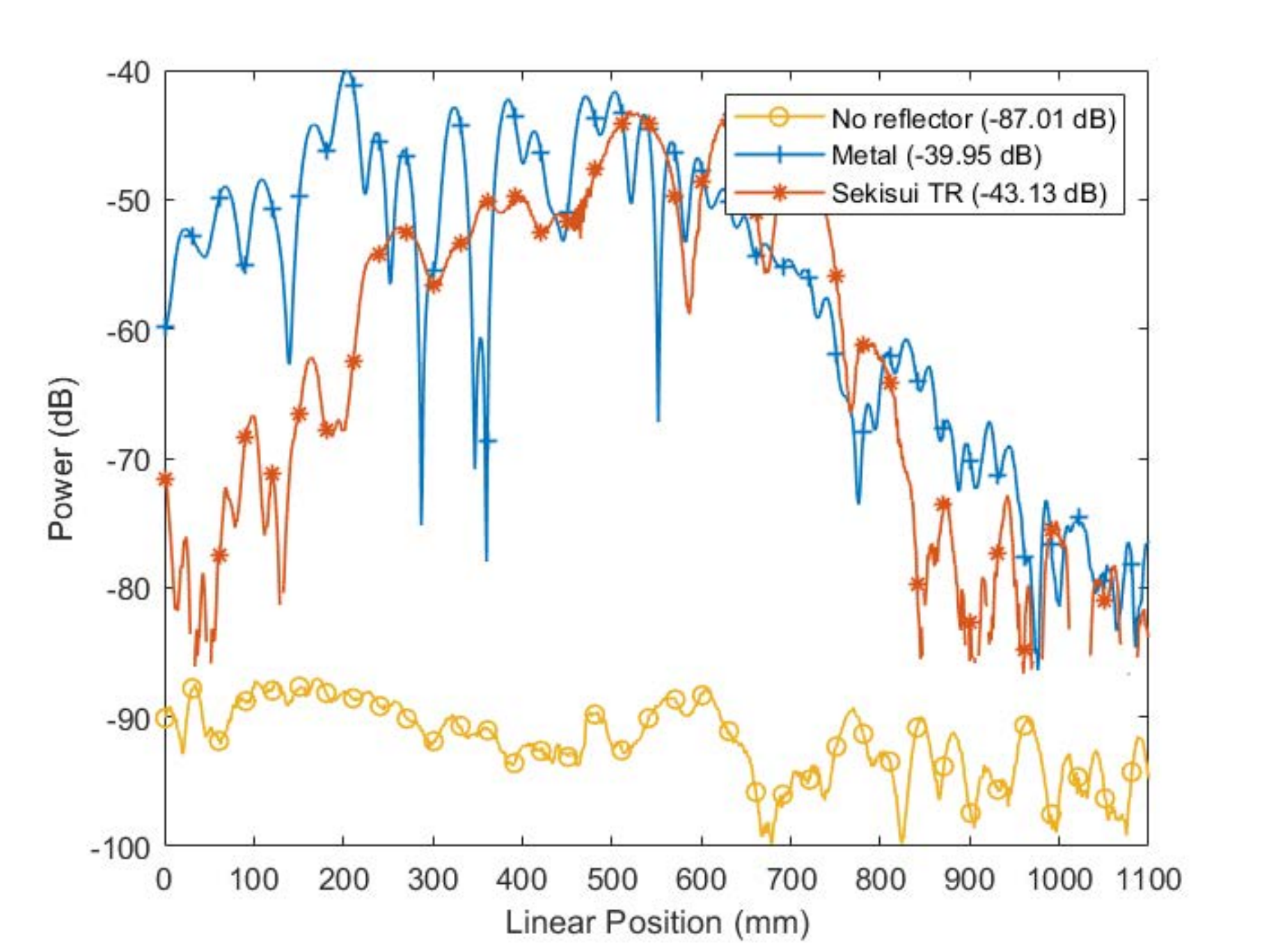}}}
        % \caption{Freq: 120 GHz, Reflector Type: Flat}
    \end{subfigure}
    \begin{subfigure}[b]{0.45\textwidth}
       {\subfloat[Frequency: 120~GHz, Shape: Convex]{\includegraphics[width=\textwidth]{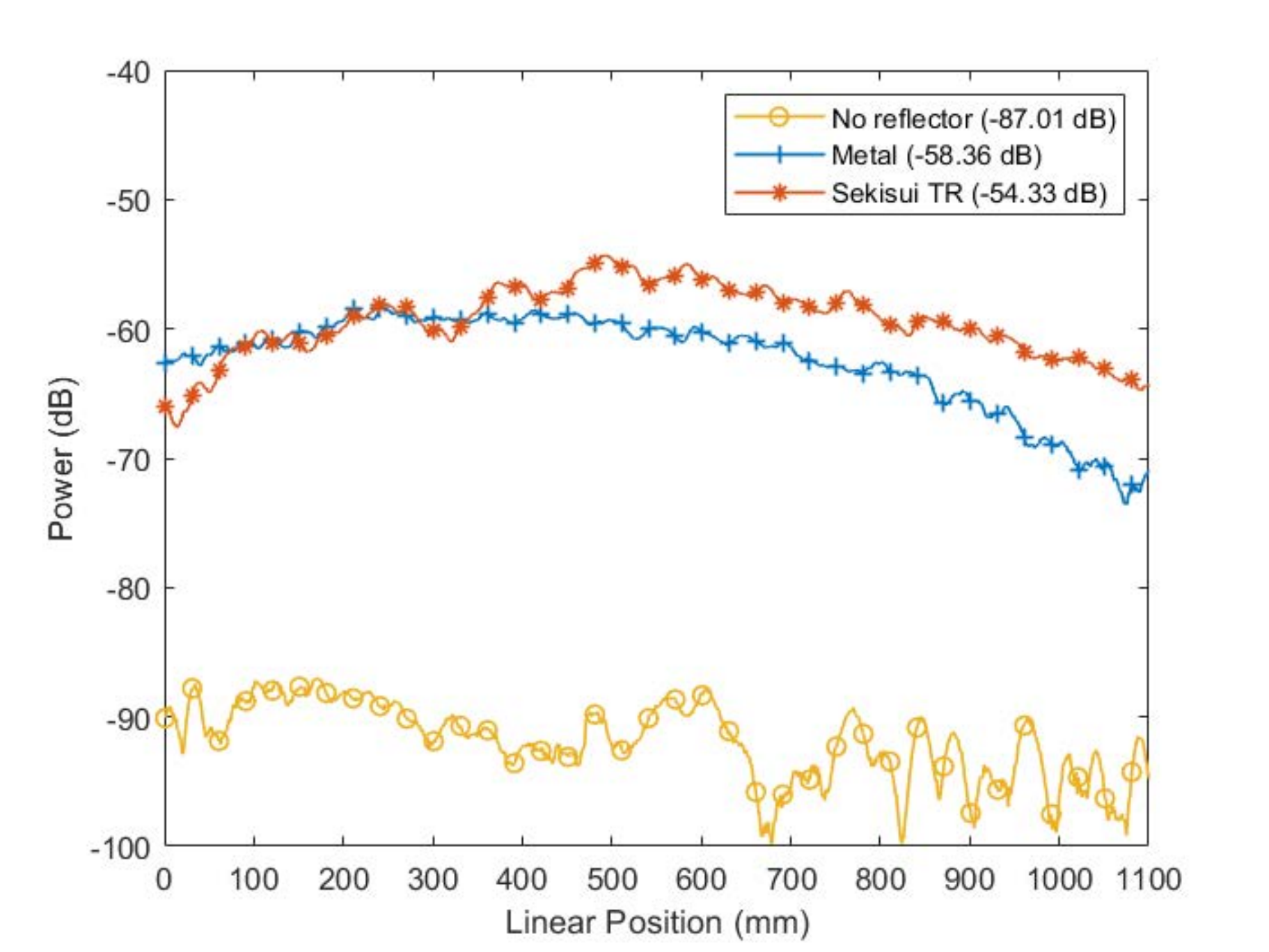}}}
       \centering
    %   \caption{Freq: 120 GHz, Reflector Type: Convex}
    \end{subfigure}
    \\
\caption{Flat and curved reflector propagation measurements at three different frequency bands.}
\label{fig:Measurement results} 
\end{figure*}

Table~\ref{tab:reflected_powers} provides a summary of the highest reflected powers across 28~GHz, 39~GHz, and 120~GHz, using flat and convex-shaped reflectors. Fig.~\ref{fig:Measurement results} presents the reflected power versus linear RX position using the metallic and Sekisui transparent reflector (TR). We compared the flat and curved reflector results for maximum power received in three different scenarios, using metallic, Sekisui TR, and no reflector. Fig.~\ref{fig:Measurement results}a represents the received power using flat-shaped reflectors at 28~GHz. This plot shows that the peak received power using no reflector, metallic reflector, and Sekisui TR are -87.01~dB, -54~dB, -49.61~dB, respectively. These values indicate that flat Sekisui TR outperforms the metallic reflector by 1~dB in its reflection capacity at 28~GHz. The power distribution curve of Sekisui TR is smoother compared to metallic reflectors. In the case of the metallic reflector, we see, large varying fringing dips. Fig.~\ref{fig:Measurement results}b displays the profile of the convex reflector at 28~GHz, where reflected power distribution is now more uniformly flat and reduced by nearly 20~dB. In this case, the peak reflected power using convex Sekisui TR is -71.40~dB and the metallic reflector is -74.69~dB. The fringing properties also seem diminished in the case of convex reflectors compared to flat reflectors. While the peak power for the convex reflector is lower than the flat reflector, coverage is more uniform and can be better at certain receiver locations.

\begin{table}[t]
    \centering
    \renewcommand{\arraystretch}{1.15}
    \begin{tabular}{|c|c|c|c|c|}\hline
     \textbf{Reflector Shape} & \textbf{Type} & \textbf{28 GHz} & \textbf{39 GHz} & \textbf{120 GHz}  \\ \hline 
    \multirow{2}{*}{\textbf{Flat}} & Sekisui & -49.61~dB & -47.36~dB & -43.13~dB\\ 
          & Metal & -54.00~dB & -55.36~dB & -39.95~dB \\ \hline
    \multirow{2}{*}{\textbf{Convex}} & Sekisui & -71.40~dB  & -78.35~dB & -54.33~dB\\ 
           & Metal &-74.69~dB & -76.72~dB & -58.36~dB \\ \hline
    \end{tabular}
    \caption{Maximum reflected power measurements using the channel sounder.}
    \label{tab:reflected_powers}
\end{table}

Fig.~\ref{fig:Measurement results}c and Fig.~\ref{fig:Measurement results}d illustrate the reflectivity properties at 39~GHz, which indicate that we receive 8~dB higher peak power using flat Sekisui TR when compared with the metallic reflector. At 39~GHz we notice around 10~dB enhancement in received power using convex-shaped reflectors compared to the no-reflector scenario. Convex Sekisui TR and convex metallic reflectors have similar performance with the metallic reflector providing around 2~dB higher received peak power.

Fig.~\ref{fig:Measurement results}e and Fig.~\ref{fig:Measurement results}f denote the reflected power at 120~GHz, using flat and convex reflectors, respectively. In the case of convex Sekisui TR, it follows the same trend as 28~GHz, and there is 4~dB increase in peak received power compared to convex metallic reflectors. However, in the case of flat reflectors, Sekisui TR provides 3~dB lower peak received power than metallic reflectors. This can be due to relatively rough surfaces at higher frequencies (above 100~GHz), which causes MPCs to undergo higher absorption and reflection loss compared to 28~GHz. Another possible explanation can be due to slight variations in the placement of the reflector, leading to higher path-loss and a narrower reflected profile at sub-THz frequency. 
%Section~\ref{Sec:4}.B provides further analysis on these measurement results.
%reason out why?
%Flat Metal Reflector ray-based results 

\subsection{Comparison of Ray Tracing and Measurement Results}\label{Sec:5}

\begin{figure*}[h!]
\centering 
    \begin{subfigure}[b]{0.3\textwidth}
    \centering 
    {\subfloat[Frequency: 28 GHz ]{\includegraphics[width=\textwidth]{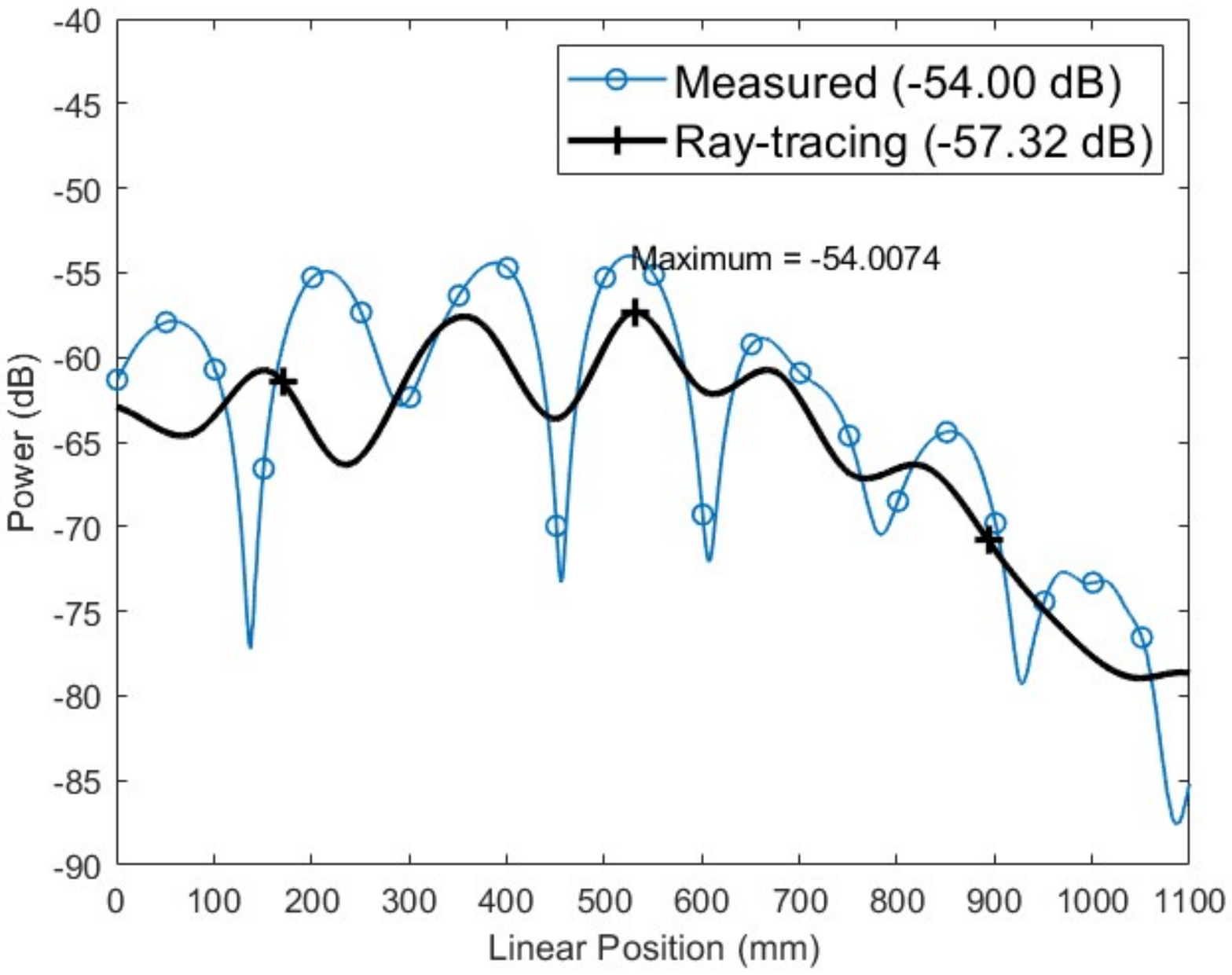}}}
    \end{subfigure}
    \begin{subfigure}[b]{0.3\textwidth}
    \centering
    {\subfloat[Frequency: 39 GHz ]{\includegraphics[width=\textwidth]{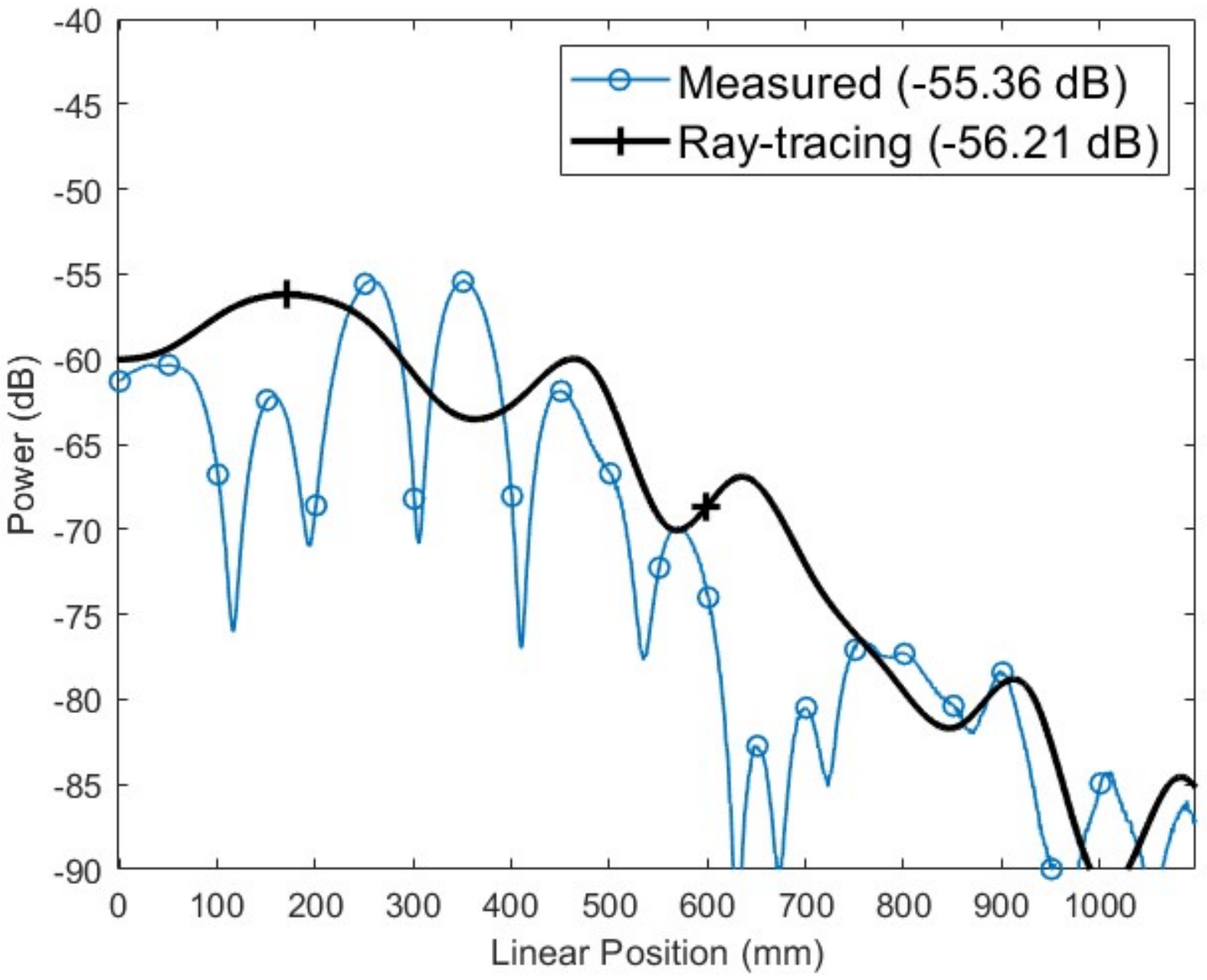}}}
    \end{subfigure}
    \begin{subfigure}[b]{0.3\textwidth}
    \centering
    {\subfloat[Frequency: 120 GHz ]{\includegraphics[width=\textwidth]{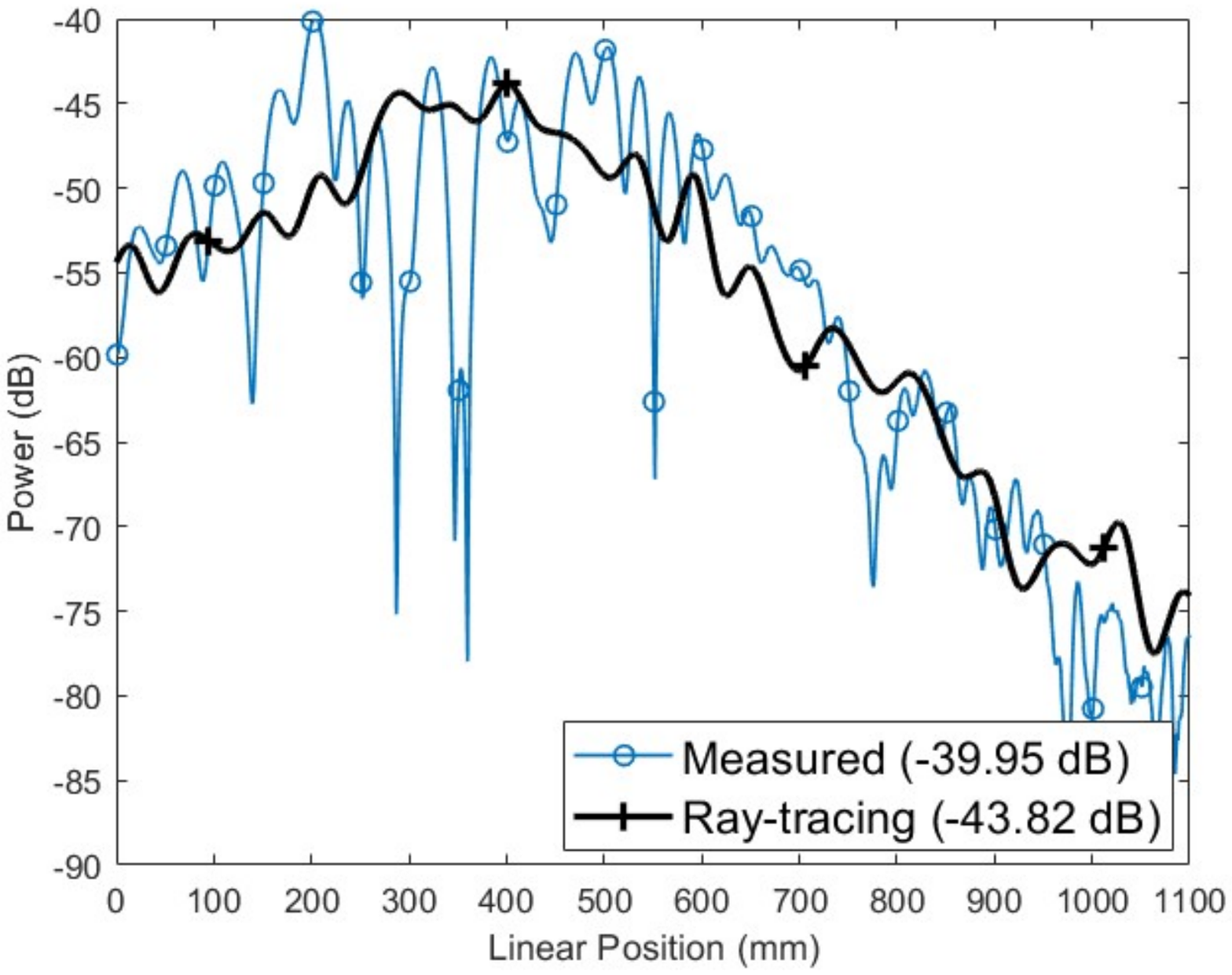}}}    
    \end{subfigure}
    \caption{{Comparison of ray-tracing simulation and measurement results (reflector type:  flat - metal reflector).}}
        \label{fig:Flat SimvsMeasured}
        \vspace{-0.65cm}\quad
\end{figure*}

The ray-based simulations are carried out in Matlab. The geometry, size of the reflector, transmit power, and antenna radiation pattern from the measurement scenario in Fig.~\ref{fig:Testscenario} are used for the simulation setup. Fig.~\ref{fig:Flat SimvsMeasured} provides the simulation results for the flat metallic reflector of size $16"\times 16"$ at $28$~GHz, $39$~GHz, and $120$~GHz. The fluctuations in the received power observed in these plots are due to constructive and destructive (CD) interference of rays from different facets~(see (\ref{Eq:RX_pwr_flat})). The number of facets used in simulations for $28$~GHz and $39$~GHz is $36$, whereas for the $120$~GHz the number of facets is $256$. A large number of facets helps to capture small power fluctuations at $120$~GHz as observed during the measurements. 

From Fig.~\ref{fig:Flat SimvsMeasured}, it can be observed that the width of the fluctuation cycles due to CD interference decreases significantly at $120$~GHz. Furthermore, as expected, the received power is highest near the RX position where the incident and reflected ray angles are the same. It is also observed that the received power decreases as we move toward the right hand side (RHS) of the linear RX positioner. The decrease is due to an increase in the azimuth angle of the rays from the main reflection position which is $30^{\circ}$ on the left hand side (LHS) from the center. The large angle of the rays in the azimuth plane results in small antenna gain and hence reduced received power~(see (\ref{Eq:RX_pwr_flat})). Overall, the simulation results for the flat metal reflector match closely with the measurement results. It may be possible to improve the match between measurement and ray tracing results using a larger number of facets for the reflector, and  using finer resolution beam patterns at the TX and the RX. %Compared to the measurements, the fades in the received power are not deep. The decrease in the received power is smaller on the LHS of the linear positioner compared to RHS because the reflected energy is mainly distributed towards the LHS.
%Curved Metal Reflectors ray-based analysis
\graphicspath{ {./images/} }

\begin{figure*}[]
\centering 
    \begin{subfigure}[b]{0.3\textwidth}
    \centering
    {\subfloat[Frequency: 28 GHz ]{\includegraphics[width=\textwidth]{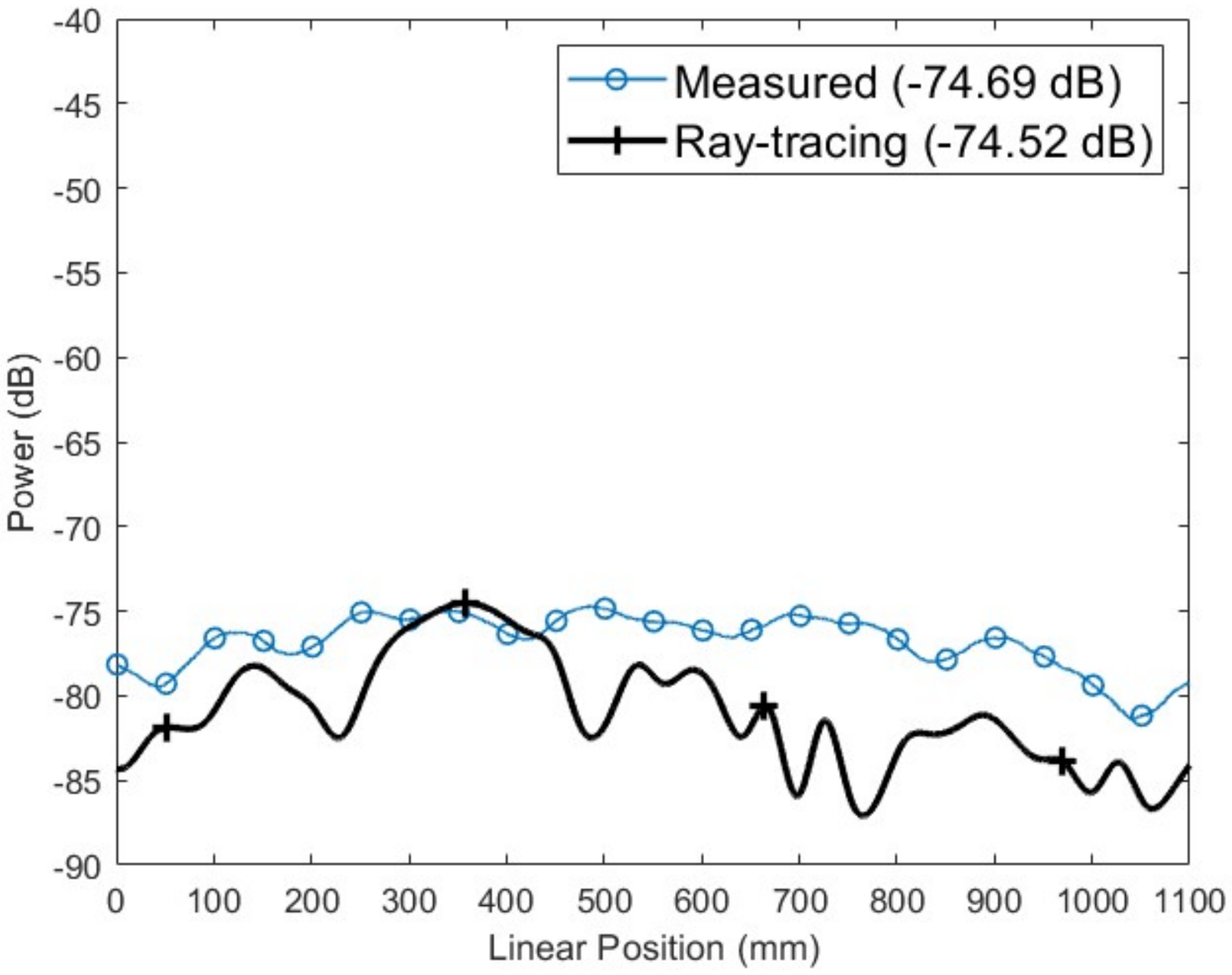}}}
    \end{subfigure}
    \begin{subfigure}[b]{0.3\textwidth}
    \centering
    {\subfloat[Frequency: 39 GHz ]{\includegraphics[width=\textwidth]{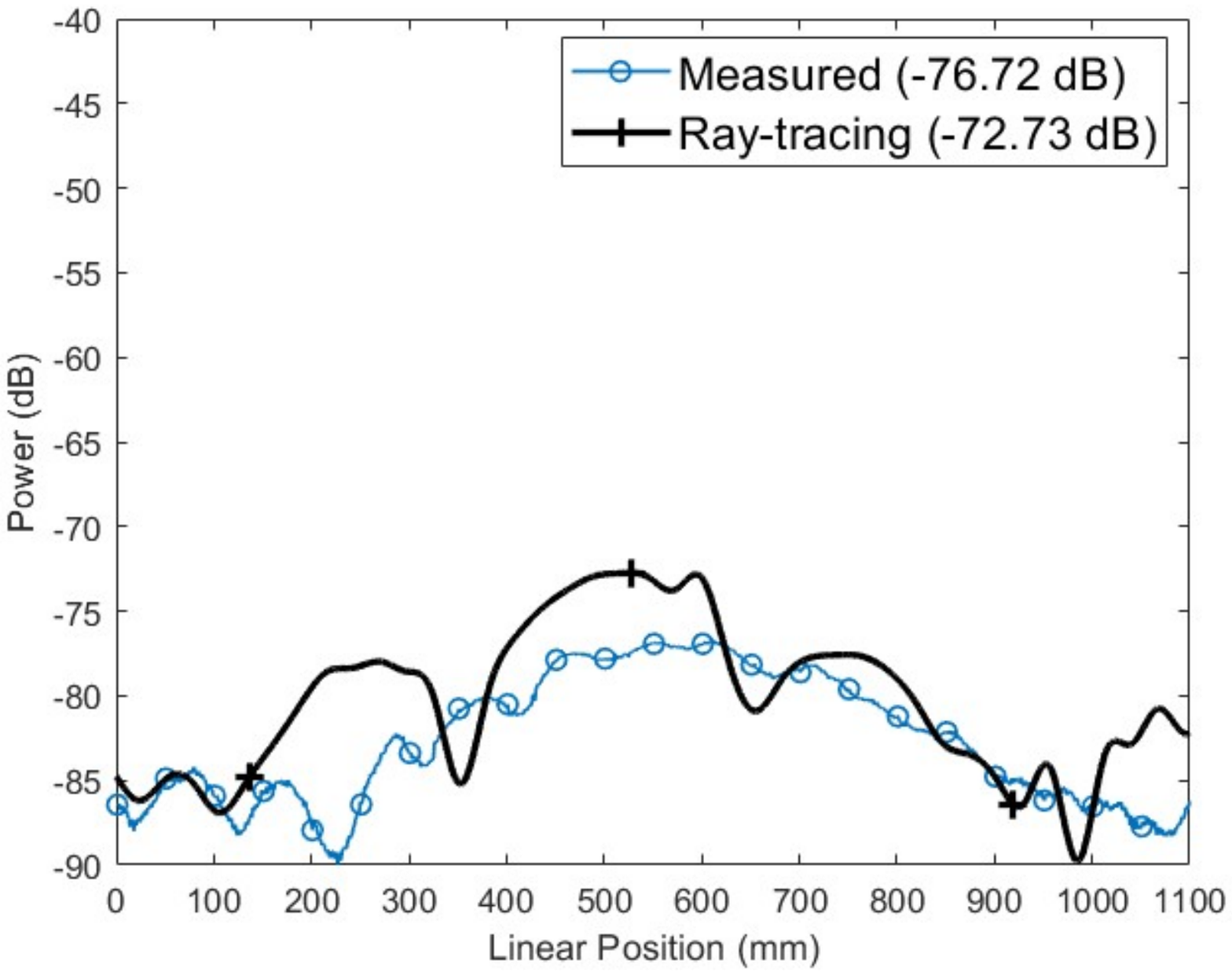}}}
    \end{subfigure}
    \begin{subfigure}[b]{0.3\textwidth}
    \centering
    {\subfloat[Frequency: 120 GHz ]{\includegraphics[width=\textwidth]{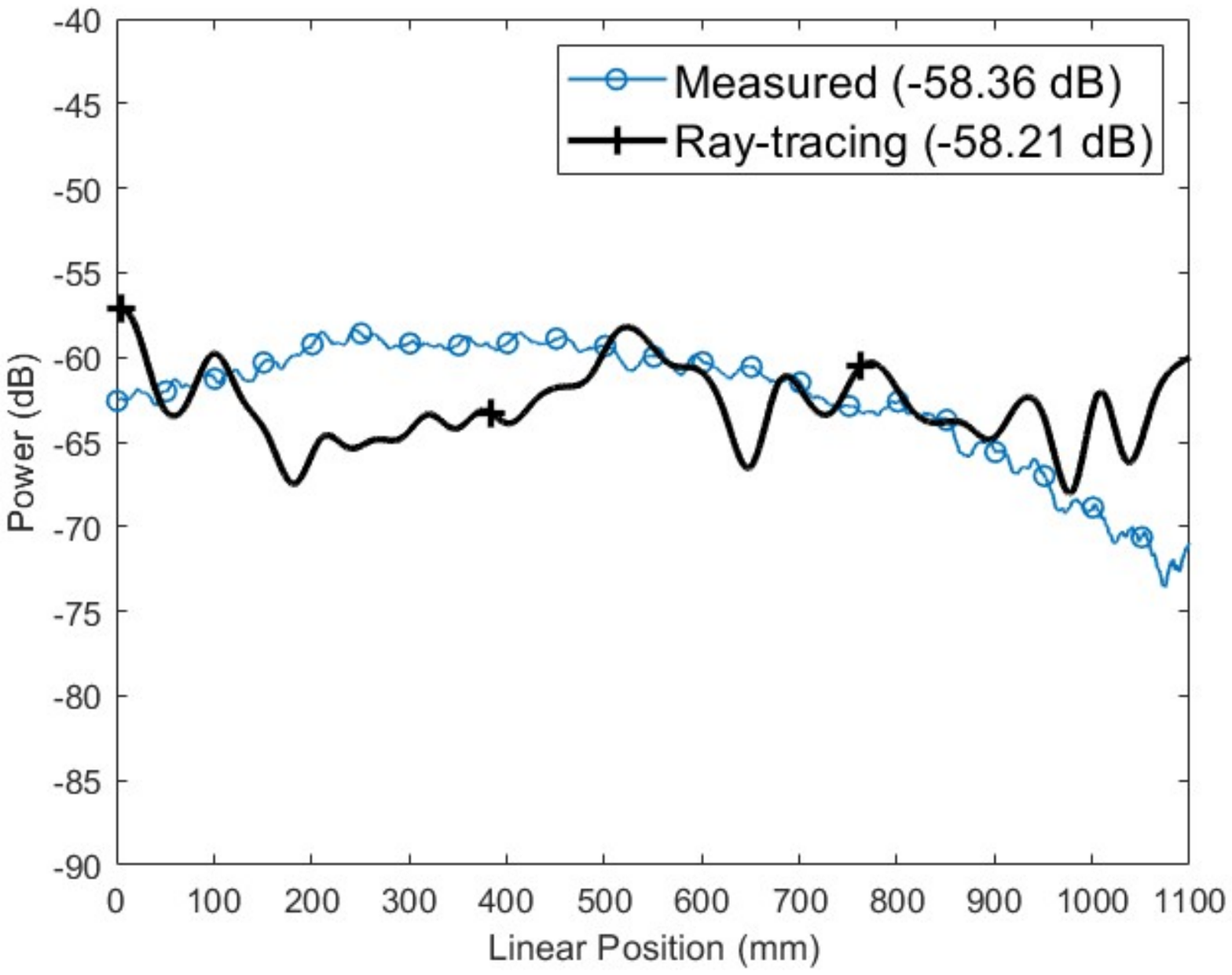}}}  
    \end{subfigure}
    \caption{{Comparison of ray-tracing simulation and measurement results (reflector type: convex - metal reflector).}}
        \label{fig:Convex SimvsMeasured}
        \vspace{-0.2cm}
\end{figure*}

Fig.~\ref{fig:Convex SimvsMeasured} displays the simulation results for the convex metal reflector of size $16"\times 16"$ at $28$~GHz, $39$~GHz, and $120$~GHz. Compared to the flat reflector, the received power does not change significantly across the RHS of the linear positioner. The small reduction of the received power on the RHS of the linear positioner compared to the flat reflector is due to uniform divergence of rays in the azimuth plane (ref Fig.~\ref{Fig:simulation_scenario_convex}). Furthermore, the overall received power is smaller for the convex reflector compared to the flat reflector. This is due to the higher attenuation of rays for the convex reflector~(\ref{Eq:RX_pwr_curve}). 
Similar to the flat reflector, we observe fluctuations in the received power for the convex reflector due to CD interference of the rays~(\ref{Eq:RX_pwr_curve}). Noticeably, small-scale fluctuations are observed at $120$~GHz flat reflector case are also observed here. %However, the fluctuations are shallow compared to flat reflector. from curved azimuth plane at different height sections~(\ref{Eq:RX_pwr_curve}). 

\section{Concluding Remarks}\label{Sec:5}
In this work, we analyze the coverage enhancements using passive reflectors at mmWave and sub-THz frequencies in indoor NLOS environments.  Channel measurements are performed using passive metallic and transparent reflectors of different shapes to understand their power profile in our propagation channel. The results reveal that flat transparent reflectors give better coverage in the NLOS region than metal reflectors in 28~GHz and 39~GHz and closely match metallic reflector performance at 120~GHz (by 3~dB). Using convex-shaped reflectors, Sekisui TR performed better than the metallic reflector at 28~GHz and 120~GHz, and slightly lower (by 2~dB) than the metallic reflector at 39~GHz in terms of the peak received power. The measurement findings for metallic reflectors were compared with ray tracing simulations in  the same environment, which correlate closely to our measured results. This suggests Sekisui TR as a viable option for passively increasing coverage in an indoor NLOS scenario. The findings also suggest that enhancements due to using passive reflectors in sub-THz bands are comparable to those in mmWave frequencies. %It can be promising for future applications, as there are fewer reflections in LOS-directed channels at these frequencies due to narrower beam antennas on both ends of the link. %At 120 GHz, the transparent reflector's performance is at its peak, indicating that these reflectors may be utilized to improve wave guiding at greater sub-THz frequencies.

\bibliographystyle{IEEEtran}
\bibliography{IEEEabrv,references}
\vspace{12pt}
\color{red}
%IEEE conference templates contain guidance text for composing and formatting conference papers. Please ensure that all template text is removed from your conference paper prior to submission to the conference. Failure to remove the template text from your paper may result in your paper not being published.

\end{document}